\documentclass[a4paper,10pt,twoside]{article} \usepackage{a4}
\usepackage[english]{babel}
\usepackage{makeidx}
\usepackage{epsfig}
\usepackage{graphicx}
\usepackage{dcolumn}
\usepackage{amsmath}
\usepackage{amssymb}
\usepackage{overcite}
\begin{document}


\newcommand{\den}{\rho}
\newcommand{\delt}{d}
\newcommand{\dist}{t}
\newcommand{\aaa}{\"a}
\newcommand{\ooo}{\"o}
\newcommand{\uuu}{\"u}
\newcommand{\vex}{v_{ext}}
\newcommand{\orb}{\varphi}
\newcommand{\nbar}{\overline{\den}}
\newcommand{\cco}{\alpha}
\newcommand{\sca}{\lambda}
\newcommand{\kF}{\kappa}
\newcommand{\F}{{\mbox{\it\footnotesize APP}}}
\newcommand{\Gr}{{\mathit{\Gamma}}}
\newcommand{\De}{{\mathit{\Delta}}}
\newcommand{\Heav}{{\mathit{\Theta}}}
\newcommand{\Lam}{{\mathit{\Lambda}}}
\newcommand{\Pot}{{\mathit{\Phi}}}
\newcommand{\gr}{\gamma}
\newcommand{\lts}{{{} \atop \textstyle\approx}\!\!\!\!\!\!{\textstyle<\atop {}}}
\newcommand{\secInt}{1}
\newcommand{\secPD}{2}
\newcommand{\secJEL}{3}
\newcommand{\secJELTWO}{4}
\newcommand{\secOV}{5}
\newcommand{\secSUM}{6}

\newcommand{\bmath}{\begin{eqnarray}}
\newcommand{\emath}{\end{eqnarray}}

\renewcommand{\topfraction}{1.0}
\renewcommand{\bottomfraction}{1.0}
\renewcommand{\textfraction}{0.0}
\renewcommand{\floatpagefraction}{1.0}
\setcounter{topnumber}{2}
\setcounter{bottomnumber}{2}
\setcounter{totalnumber}{3}
\itemindent2.0cm
\newcommand{\samefootnotemark}{\addtocounter{footnote}{-1} \footnotemark}
\newcommand{\setfootnotemark}[1]{\addtocounter{footnote}{#1} \footnotemark}

\renewcommand{\thefootnote}{\fnsymbol{footnote}}


\def\hf{{\rm HF}}
\def\uu{\uparrow\uparrow}
\def\ud{\uparrow\downarrow}
\def\du{\downarrow\uparrow}
\def\dd{\downarrow\downarrow}
\def\rv{{\bf r}}
\def\Rv{{\bf R}}
\def\kv{{\bf k}}
\def\Kv{{\bf K}}
\def\e{{\rm e}}
\def\be{\begin{equation}}
\def\ee{\end{equation}}
\def\bea{\begin{eqnarray}}
\def\eea{\end{eqnarray}}
\thispagestyle{empty}

\vspace*{4cm}
\begin{flushright}
{\Large\bf UNIFORM ELECTRON GAS FROM}
\end{flushright}

\vspace{-0.7cm}
\begin{flushright}
{\Large\bf TWO-PARTICLE WAVEFUNCTIONS}
\end{flushright}

\vspace{0.5cm}
\begin{flushleft}
{\large Paola Gori-Giorgi
\footnote{Paola Gori-Giorgi, Unit\`a INFM and Department of Physics, 
University of Rome
``La Sapienza'', Piazzale A. Moro 2, I-00185 Rome, Italy. E-mail:
{\tt gp.giorgi@caspur.it}}}
\end{flushleft}

\vspace{0.5cm}
\noindent
{\bf\secInt . INTRODUCTION}
\vspace{0.3cm}

A common approach to the
many-electron problem (atoms, molecules and solids) is its
transformation into a fewer-particle problem. In Density Functional
Theory\cite{HK,KS} (DFT)
one rewrites the ground-state energy of a many-electron
system as a functional of just the one-electron density $n(\rv)$, the
diagonal part of the one-particle reduced density matrix.
In recent years, much attention has been devoted to approaches which
rewrite the system energy as a functional of the one- or the two-particle 
reduced density 
matrix.\cite{mazziotti,Yasuda,GS,GU,ziescheCreta,ziescheRDM,pernJCP} 
With respect to DFT, these approaches have the advantage that the kinetic
energy functional is known and that they provide more information about the
many-body wavefunction. 
In this contest, the relevance of two-electron
wavefunctions (geminals) has been pointed out:
the use of an antisymmetrized geminal 
power wavefunction, i.e., a many-body wavefunction built in terms of geminals
with the correct symmetry under particle permutation,
provides the basis for a formally correct one-particle reduced density
matrix functional theory.\cite{mazziotti}

With these concepts in mind, it is of great interest to start with a simple 
many-body system, the uniform electron gas, a
limit which should be recovered
by any approximate approach to the many-electron problem of nonuniform
density. 
A more detailed discussion about reduced-density matrix functionals applied
to the uniform electron gas can be found in
Refs.~\citen{ziescheCreta,pernal,mioref,zieschequi}. 
In this paper, an approximation for the unknown two-electron
wavefunctions (geminals) of the uniform electron gas is found, starting from
the effective screened Coulomb potential proposed by 
Overhauser.\cite{Overhauser}
The short-range part (small electron-electron distances) of the 
corresponding pair 
density is found to be in accurate agreement with the newest Quantum Monte
Carlo data.\cite{OHB} 
This means that the short-range part of these geminals is reliable and
could serve as a useful term of comparison for 
other two-electron approaches to the uniform electron 
gas.\cite{ziescheCreta,pernal,mioref,zieschequi}

\vspace{1cm}
\noindent
{\bf\secPD . ELECTRON PAIR-DENSITIES, ETC.}
\vspace{0.2cm}

Given an $N$-electron wavefunction $\Psi(\rv_1\sigma_1,...\rv_N\sigma_N)$, 
we define the pair density
\begin{equation}
\rho_2(\rv,\rv') =  N(N-1) 
 \sum_{\sigma_i=1}^N\
 \int \prod_{j=3}^N d\rv_j
|\Psi(\rv\sigma_1,\rv'\sigma_2,...\rv_N\sigma_N)|^2,
\label{eq_rho2}
\end{equation} 
the one-particle density matrix 
\begin{equation}
\rho_1(\rv,\rv') =  N\sum_{\sigma_i=1}^N\int
\Psi^*(\rv\sigma_1,\rv_2\sigma_2,...\rv_N\sigma_N)
 \Psi(\rv'\sigma_1,\rv_2\sigma_2,...\rv_N\sigma_N)
d\rv_2...d\rv_N,
\label{eq_rho1}
\end{equation}
and the one-electron density
\begin{equation}
n(\rv)=\rho_1(\rv,\rv)=\frac{1}{N-1}\int d\rv'\rho_2(\rv,\rv').
\label{eq_dens}
\end{equation}
While $n(\rv)d\rv$ is the probability of finding an electron in $d\rv$,
$\rho_2(\rv,\rv')d\rv d\rv'$ is the probability of finding 
one electron in $d\rv$
and another in $d\rv'$. We also define the pair-distribution function
$g(\rv,\rv')$:
\begin{equation}
\rho_2(\rv,\rv')=n(\rv)n(\rv')g(\rv,\rv').
\label{eq_g}
\end{equation}
By integrating Eq.~(\ref{eq_rho2}) over $\rv'$, we find that
\begin{equation}
\int d\rv' n(\rv')\,[g(\rv,\rv')-1]=-1.
\label{eq_sumrule}
\end{equation}
In other words, the density $n(\rv')[g(\rv,\rv')-1]$ of the 
exchange-correlation hole around an electron at $\rv$ represents a deficit
of one electron.\par

\vspace{1cm}
\noindent
{\bf\secJEL . UNIFORM ELECTRON GAS (JELLIUM)}
\vspace{0.3cm}

The three-dimensional jellium model consists of $N$ electrons enclosed in a box
of volume $V$ (periodically repeated in space) in the
presence of a neutralizing background of uniform positive charge 
density $n^{+}=N/V$. The 
non relativistic jellium is thus governed by the hamiltonian 
(in Hartree atomic units):
\begin{equation}
H=\sum_{i=1}^N\frac{{\bf p}_i^2}{2}+\frac{1}{2}\sum_{i\ne j=1}
^N\frac{1}{|{\bf r}_i-{\bf r}_j|}+\Lambda,
\label{eq_hamjellium}
\end{equation}
where $\Lambda$ represents the effect of the
background. The hamiltonian of Eq.~(\ref{eq_hamjellium})
describes a model solid whose positive ionic charges are smeared
throughout the whole crystal volume to yield a shapeless, uniform
positive background (whence the nickname of jellium).
When studying this model, one is usually interested in its macroscopic
properties, i.e., the thermodynamic limit ($N,V\to \infty$ keeping $n=N/V$
constant) of its extensive physical quantities per particle or per 
volume.
Two numbers are enough to describe its zero-temperature phase 
diagram,
namely, the one-electron density $n=N/V$ and the spin
polarization $\zeta=|N_{\uparrow}-N_{\downarrow}|/N$,
where $N_{\uparrow(\downarrow)}$ is the number of spin-up (down)
electrons ($N=N_{\uparrow}+N_{\downarrow}$).
Instead of the particle density $n$, it is often convenient to
use the Wigner-Seitz radius $r_s$ (in units of the Bohr
radius) given by
$r_s = \left(\frac{4\pi}{3}\,n\right)^{-1/3}$;
it is also useful to define the Fermi wavevector $k_F$, simply
related to the $r_s$ parameter by
$k_F=\left(3\pi^2n\right)^{1/3}=
(\frac{9\pi}{4})^{1/3}\frac{1}{r_s}$.

Beeing the system homogeneous and isotropic, the 
pair-distribution function only depends on $r=|\rv_1-\rv_2|$, and
parametrically on $r_s$ and $\zeta$.


\vspace{1cm}
\noindent
{\bf\secJELTWO . JELLIUM FROM TWO-ELECTRON WAVEFUNCTIONS}
\vspace{0.3cm}

The pair-distribution function of jellium can be built starting
from two-electron wavefunctions.\cite{Overhauser,GGP}
We first rewrite the non-interacting
gas (ideal Fermi gas) in terms of two-electron wavefunctions. This first step
can seem redundant, since the ideal gas can be written in terms of 
one-electron wavefunctions, as a Slater determinant of plane waves. However,
treating first the ideal gas is essential to set the proper normalization.
The two-electron wavefunctions for the interacting gas, in fact, are
found by solving a scattering problem in an effective potential,
with the normalization condition that the ideal gas is recovered when
the potential is set to zero. In this paper, only the $\zeta=0$
case is analysed. For the generalization to the $\zeta \neq 0$ gas
see Ref.~\citen{GGP}. 

\vspace{0.3cm}
\noindent
{\bf\secJELTWO .1. Ideal Fermi gas}
\vspace{0.3cm}

\begin{figure}
\begin{center}
\includegraphics[width=8cm]{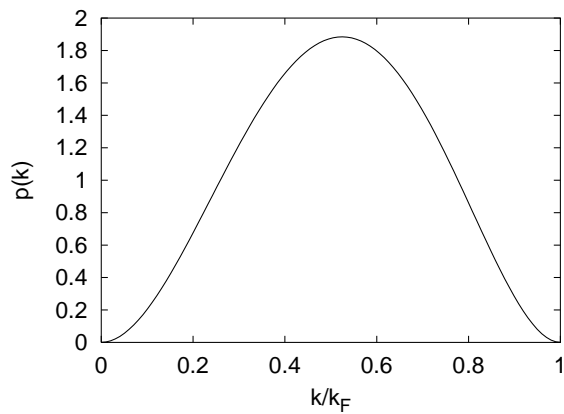}
\end{center} 
\caption{Relative-momentum ($k=\frac{1}{2}|\kv_2-\kv_1|$) probability 
distribution $p(k)$ for the noninteracting gas.}
\label{fig_pk}
\end{figure}

If we select a pair of electrons at random in the spin-unpolarized 
uniform gas, there is one chance in four that they will be in the singlet
state, $\ud-\du$, and three chances in four that they will be in one of the
triplet states, $\uu$, $\dd$, $\ud+\du$. In the case of no 
electron-electron interaction, the corresponding two-electron 
spatial wavefunctions can be rewritten in the center-of-mass reference
system as
\begin{equation}
\Psi(\rv,\Rv)=\frac{1}{\sqrt{2}}\,\e^{i\,\Kv\cdot\Rv}\,\left(
\e^{i\,\kv\cdot\rv}\pm\e^{-i\,\kv\cdot\rv}\right),
\label{eq_freewave}
\end{equation}
where ``$+$''is for the singlet state and ``$-$'' is for
the triplet state, and
\begin{equation}
 \Rv=\tfrac{1}{2}(\rv_1+\rv_2), \qquad  \rv=\rv_2-\rv_1, \qquad
 \Kv=\kv_1+\kv_2, \qquad \kv=\tfrac{1}{2}(\kv_2-\kv_1).
\end{equation}
Beeing the system isotropic, it is convenient to
expand the plane waves into spherical harmonics
\begin{equation}
\e^{i\,\kv\cdot\rv}=\sum_{\ell=0}^{\infty}(2\ell+1)\,i^{\ell}\,P_{\ell}(\cos
\theta)\,j_{\ell}(k\,r),
\end{equation}
where $P_{\ell}$ are Legendre polynomials and $j_{\ell}$ are spherical Bessel
functions. Then
\begin{eqnarray}
\Psi_{\mathrm{singlet}}(\rv,\Rv) & = & \sqrt{2}\,\e^{i\,\Kv\cdot\Rv}
\sum_{\stackrel{\ell=0}{\mathrm{even}\; \ell}}^{\infty}(2\ell+1)\,i^{\ell}\,
P_{\ell}(\cos\theta)\,j_{\ell}(k\,r) \label{eq_psising} \\
\Psi_{\mathrm{triplet}}(\rv,\Rv) & = & \sqrt{2}\,\e^{i\,\Kv\cdot\Rv}
\sum_{\stackrel{\ell=1}{\mathrm{odd}\; \ell}}^{\infty}(2\ell+1)\,i^{\ell}\,
P_{\ell}(\cos\theta)\,j_{\ell}(k\,r). \label{eq_psitrip}
\end{eqnarray}
The pair-distribution function can be obtained by $\Psi_{\mathrm{singlet}}$
and $\Psi_{\mathrm{triplet}}$ by giving them the proper weight,
\begin{equation}
g(r)=\tfrac{1}{4}\langle |\Psi_{\mathrm{singlet}}(\rv)|^2 \rangle
+\tfrac{3}{4}\langle |\Psi_{\mathrm{triplet}}(\rv)|^2 \rangle,
\label{g_frompsi}
\end{equation}
and by considering that in the uniform electron gas there is a probability
$p(k)$ that two electrons have a given relative momentum $k=\frac{1}{2}|\kv_1
-\kv_2|$. In Eq.~(\ref{g_frompsi}) the symbol
$\langle\rangle$ means, in fact,  that an average over $p(k)$ and over the solid angle
has to be performed.
If one is interested in the 
spin-resolved pair-distribution functions,
$g_{\uu}(r)$ and $g_{\ud}(r)$, corresponding to parallel- and
antiparallel-spin interactions, and such that for the unpolarized
gas 
\begin{equation}
g=\tfrac{1}{2}(g_{\uu}+g_{\ud}), 
\end{equation}
one has to consider that $\frac{1}{3}$ of the triplet state 
($\ud+\du$) contributes to the antiparallel-spin 
correlations and $\frac{2}{3}$ of it ($\uu$ and $\dd$) to the 
parallel-spin correlations. 
So, we have
\begin{eqnarray}
g_{\ud}(r) & = & \tfrac{1}{2}\langle |\Psi_{\mathrm{singlet}}(\rv)|^2 \rangle
+\tfrac{1}{2}\langle |\Psi_{\mathrm{triplet}}(\rv)|^2 \rangle\\
g_{\uu}(r) & = & \langle |\Psi_{\mathrm{triplet}}(\rv)|^2\rangle,
\end{eqnarray}
where $\langle \rangle$ denotes again average over $p(k)$ and over the solid angle.
Performing the spherical average over the solid angle, we obtain:
\begin{eqnarray}
g_{\ud}(r) & = & \sum_{\ell=0}^{\infty}(2\ell+1)\langle 
j_{\ell}^2(k\,r)\rangle
\label{eq_gudnonint} \\
g_{\uu}(r) & = & 2\sum_{\stackrel{\ell=1}{\mathrm{odd}\; \ell}}^{\infty}
(2\ell+1)\langle j_{\ell}^2(k\,r)\rangle.
\label{eq_guunonint}
\end{eqnarray}
Equation~(\ref{eq_gudnonint}) immediately gives the exact result for a 
noninteracting gas, i.e., $g_{\ud}(r)=1$ for each $r$. To obtain the 
noninteracting $g_{\uu}(r)$ from Eq.~(\ref{eq_guunonint}), we need
to average over $k$. In the noninteracting electron gas,
the probability distribution $p(k)$ for 
$k=\tfrac{1}{2}|\kv_2-\kv_1|$ can
be obtained geometrically by considering two three-dimensional 
vectors $\kv_1$ and $\kv_2$ with $0\le |\kv_{1(2)}| \le k_F$, where
$k_F$ is the Fermi wavevector.
The probability $p(k)$ is then
\begin{equation}
p(k)=24\frac{k^2}{k_F^3}-36\frac{k^3}{k_F^4}+12\frac{k^5}{k_F^6},
\label{eq_pk}
\end{equation}
with $k$ ranging from $0$ to $k_F$ (see Fig.~\ref{fig_pk}). 
By using two known series which involve $j_{\ell}^2$,
\begin{equation}
\sum_{\ell=0}^{\infty}(2\ell+1)
j_{\ell}^2(x)=1 \qquad
\sum_{\ell=0}^{\infty}(-1)^{\ell}(2\ell+1)
j_{\ell}^2(x)=\frac{\sin(2x)}{2x},
\label{eq_s1}
\end{equation}
we can rewrite $g_{\uu}$ as
\begin{equation}
g_{\uu}(r)=\sum_{\ell=0}^{\infty}(2\ell+1)\langle 
j_{\ell}^2(k\,r)\rangle-\sum_{\ell=0}^{\infty}(-1)^{\ell}(2\ell+1)\langle 
j_{\ell}^2(k\,r)\rangle=\left\langle 1-\frac{\sin(2kr)}{2kr}\right\rangle.
\end{equation}
This means:
\begin{equation}
g_{\uu}(r)=\int_0^{k_F}\left[1-\frac{\sin(2kr)}{2kr}\right]
\left(24\frac{k^2}{k_F^3}-36\frac{k^3}{k_F^4}+12\frac{k^5}{k_F^6}\right)
\,dk.
\end{equation}
Performing this integral gives the known Hartree-Fock (noninteracting) $g_{\uu}$,
\begin{equation}
g_{\uu}^{\rm HF}=1-9\left[\frac{\sin(k_Fr)-k_Fr\cos(k_Fr)}{(k_Fr)^3}\right]^2.
\label{g_HF}
\end{equation}
Numerically, in the range $0\le k_Fr\le 6$ and with 
a truncation of the infinite sum over $\ell$ at $\ell_{\rm max}=7$,
Eq.~(\ref{eq_guunonint}) reproduces the exact Eq.~(\ref{g_HF}) 
within an accuracy of $10^{-6}$. When the scaled variable $k_Fr$ (or
equivalently $r/r_s$) is used, the pair-distribution function of the
noninteracting gas does not depend on $r_s$. The explicit dependence on $r_s$
only appears when Coulomb repulsion is taken into account in the wavefunction.

\vspace{0.3cm}
\noindent
{\bf\secJELTWO .2. Interacting electron gas}
\vspace{0.3cm}

The interacting case can now be treated by introducing an effective
potential which describes the electron-electron interactions in a uniform
electron gas. The spherical Bessel functions $j_{\ell}$ which appear in
Eqs.~(\ref{eq_gudnonint}) and~(\ref{eq_guunonint}) are solution of the
noninteracting radial Schr\"odinger equation 
\begin{equation}
\left[\frac{d^2}{dr^2}-\frac{\ell(\ell+1)}{r^2}+k^2
\right]u_{\ell} = 0 \qquad \qquad u_{\ell}=kr\,j_{\ell}(kr).
\label{eq_nonint}
\end{equation}
If we introduce an effective radial potential $V(r,r_s)$ which depends on
the electron-electron distance $r$ and parametrically on the electron density
$r_s$, we can solve
the corresponding interacting radial Schr\"odinger equation, 
\begin{equation}
\left[\frac{d^2}{dr^2}-\frac{\ell(\ell+1)}{r^2}-V(r,r_s)+k^2
\right]u_{\ell} = 0 \qquad \qquad u_{\ell}=kr\,R_{\ell}(r,k,r_s).
\label{eq_int}
\end{equation}
The radial functions $R_{\ell}$ will depend parametrically on $k$ and on $r_s$.
We can insert them into Eqs.~(\ref{eq_gudnonint}) and~(\ref{eq_guunonint}), and
find the corresponding $g_{\ud}$ and $g_{\uu}$:
\begin{eqnarray}
g_{\ud}(r,r_s) & = & \sum_{\ell=0}^{\infty}(2\ell+1)\langle 
R_{\ell}^2(r,k,r_s)\rangle
\label{eq_gudint} \\
g_{\uu}(r,r_s) & = & 2\sum_{\stackrel{\ell=1}{\mathrm{odd}\; \ell}}^{\infty}
(2\ell+1)\langle R_{\ell}^2(r,k,r_s)\rangle.
\label{eq_guuint}
\end{eqnarray}
To compute the average over all the possible relative $k$ (represented again
by the symbol $\langle\rangle$) one should in principle
use the interacting momentum distribution, which deviates from the Fermi
step function of the ideal gas. This would slightly change $p(k)$ 
of Fig.~\ref{fig_pk}
by adding a ``tail'' for $k>k_F$ and lowering the maximum. 
For practical purposes, the use of
the noninteracting $p(k)$ is enough to give good results.\cite{GGP} 
Notice that unless the potential $V(r,r_s)$ is very sophisticated, the
treatment just described
 will fail to recover the long-range correlations, which are mainly
governed by collective modes, and will fail to satisfy the 
particle-conservation sum rule of Eq.~(\ref{eq_sumrule}).

\vspace{1cm}
\noindent
{\bf\secOV . SOLUTION OF THE OVERHAUSER MODEL}
\vspace{0.3cm}

In this section, we compute an interacting pair-distribution function
following the procedure just described
by using the simple model potential $V(r,r_s)$ proposed 
by Overhauser.\cite{Overhauser} This simple model gives very good results for the
short-range ($r<r_s$) part of $g(r)$.\cite{GGP}
\vspace{0.3cm}

\noindent
{\bf\secOV .1. The Overhauser potential}
\vspace{0.3cm}

Overhauser\cite{Overhauser} proposed a simple and reasonable model
for the screened Coulomb repulsion $V(r,r_s)$ in the uniform electron gas:
he took the
sphere of volume $n^{-1}$ as the boundary within which the screening
charge density is $ne$ and outside of which it is zero. We thus have
the electrostatic potential due to a point charge $-e$ in the origin
plus a sphere around it of radius $r_s$ and of uniform 
positive charge density $ne$. In Hartree atomic units:
\begin{eqnarray}
  V(r,r_s)    = &  \frac{1}{r_s}\left(\frac{r_s}{r}+\frac{r^2}{2r_s^2}-\frac{3}{2}
\right)  \qquad  &  r\le r_s \nonumber  \\
  V(r,r_s)    =  & 0    & r>r_s. \label{eq_potOv}
\end{eqnarray}
This is equivalent to assuming that in the interacting gas 
the probability of finding three electrons in a sphere
of radius $r_s$ is exactly zero, an assumption which is nearly
true. In fact, numerical estimates of this probability
for the electron gas
show that it is indeed small.\cite{3p} (At $r_s=5$ 
the ratio between the probabilities of finding three
and two electrons in the same sphere of radius $r_s$ is about $1/11$; 
for larger $r_s$ this ratio is lower, and for smaller $r_s$ it is higher, being
about $1/7$ at $r_s=0$.) 
Thus, for interelectronic distances $r<r_s$ we expect
the Overhauser potential to be close to the true potential felt
by an electron moving in a uniform electron gas when another
electron is fixed at the origin. In the region $r>r_s$ the potential is set to
zero, and so is not expected to be reliable.
We also expect to have results that become more accurate as the density
decreases, since the probability of having three electrons in the
same sphere of radius $r_s$ becomes lower and lower.
Finally, at high and intermediate densities our results will be much closer
to the true $g(r)$ for antiparallel-spin correlations than for 
parallel-spin ones. When two electrons of opposite spins 
are in the same sphere of
radius $r_s$, a third electron is excluded from the
sphere because of both the Pauli principle and the Coulomb repulsion.
For a pair of parallel-spin electrons, only the Coulomb repulsion prevents
a third electron of opposite spin
from entering the sphere of radius $r_s$, a mechanism
which becomes less efficient as the density 
(and thus the kinetic energy) increases.

\vspace{0.3cm}
\noindent
{\bf\secOV .2. Solution of the model}
\vspace{0.3cm}

\begin{figure}
\includegraphics[width=6.45cm]{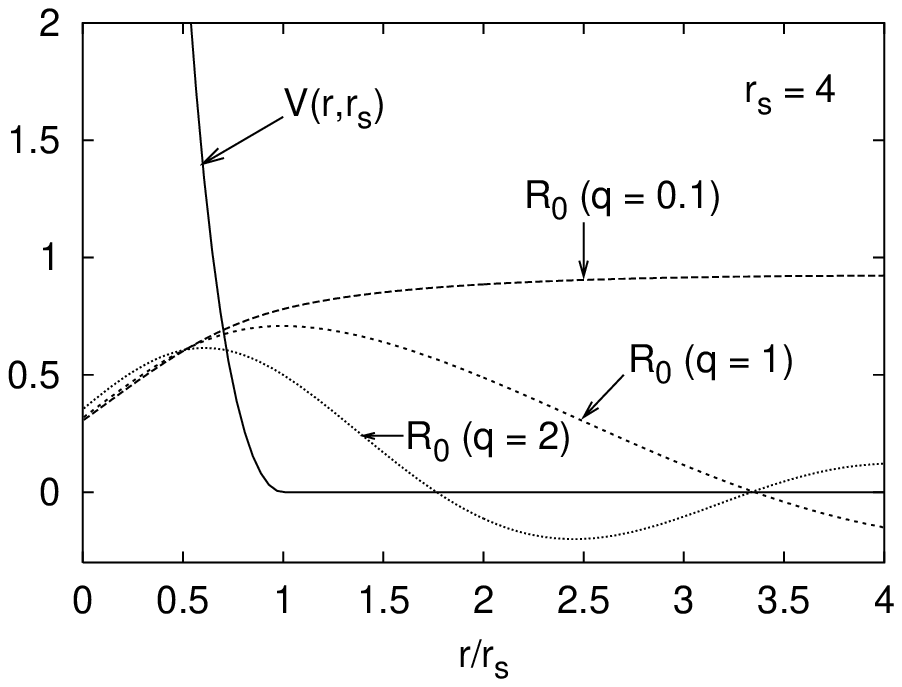}
\includegraphics[width=6.45cm]{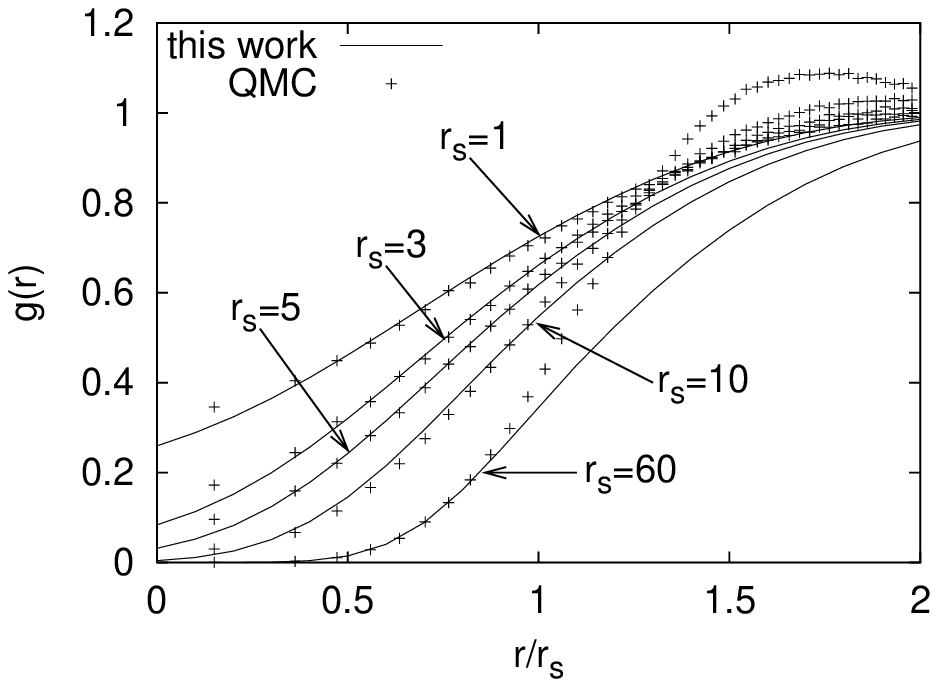}
\caption{Left panel: the Overhauser potential $V(r,r_s)$
for $r_s=4$, together with
the radial wavefunctions $R_{\ell=0}$ for different values of the
scaled relative momentum $q=k r_s$. Right panel: pair-distribution 
functions $g(r)$ of jellium computed with the Overhauser potential
compared to the Quantum Monte Carlo data of Ortiz, Harris and Ballone.\cite{OHB}}
\label{fig_R0vsqandg}
\end{figure}

The radial Schr\"odinger equation corresponding to the Overhauser potential
can be solved analytically (but not in closed-form),\cite{GGP} and the functions
$R_{\ell}(s,q,r_s)$ (where $s=r/r_s$ and $q=kr_s$ are scaled variables) are found.
As an example, in the left panel of Fig.~\ref{fig_R0vsqandg}, the radial
functions corresponding to $r_s=4$, $\ell=0$, and three different values 
of the scaled relative momentum $q=k r_s$ are reported, together
with the effective potential. 
For an unpolarized uniform gas, the probability distribution for $q$ can be
obtained from Eq.~(\ref{eq_pk}) by setting $k_F=\frac{1}{r_s}(\frac{9\pi}{4})^{1/3}$,
\begin{equation}
p(q)=\tfrac{16}{3\pi}q^2\left[2-\left(\tfrac{12}{\pi}\right)^{1/3}q
+\tfrac{4}{9\pi}\,q^3\right],
\label{eq_pq}
\end{equation}
with $q$ ranging from $0$ to $(9\pi/4)^{1/3}\approx 1.91916$. This $p(q)$
is exact for a noninteracting gas. As said, $p(q)$ for the interacting gas
slightly deviates from Eq.~(\ref{eq_pq}), and depends explicitly on $r_s$.
However, in the region where the
potential is reliable, $r/r_s\le 1$, we see from Fig.~\ref{fig_R0vsqandg}
that the $q$-dependence of $R_{\ell}$
is very weak, as already pointed out by 
Overhauser.\cite{Overhauser} We thus expect to have no significant change
in the short-range ($r<r_s$) part of $g$ if we use an interacting momentum
distribution instead of Eq.~(\ref{eq_pq}).

In the right panel of Fig.~\ref{fig_R0vsqandg} the $g(r)$ obtained by solving
the Overhauser model are reported, and compared with the newest diffusion 
Quantum Monte Carlo (QMC) data of Ortiz, Harris, and Ballone.\cite{OHB}
In the range $0\le r/r_s \le 2$, a truncation of the
infinite sum over $\ell$ in Eqs.~(\ref{eq_gudint}) and~(\ref{eq_guuint}) 
at $\ell_{\rm max}=7$ is enough to reach good convergence.\cite{GGP}
We see that there is accurate agreement with the QMC
data for $0.5\lesssim r/r_s\lesssim 1$ for a wide 
range of electron densities. In the shortest-range region,
 $r/r_s \lesssim 0.5$, the QMC data
are known to suffer large errors and are not so reliable.
(In this region there is in fact a significant discrepancy between the
data from Ref.~\citen{OHB} and those from Ref.~\citen{cepald}.)
Thus, for $r/r_s \lesssim 0.5$ the present treatment 
should provide results much closer to the true $g(r)$. As said, for $r/r_s>1$
the results obtained with the Overhauser potential are not reliable.

\vspace{1cm}
\noindent
{\bf\secSUM . SUMMARY AND CONCLUSIONS}
\vspace{0.3cm}

Two-electron wavefunctions (geminals) for the uniform electron gas 
of density $n=3/4\pi r_s^3$ are
found using the simple screened Coulomb potential proposed by
Overhauser. These wavefunctions give pair-densities in agreement
with the QMC simulations in the short-range ($r<r_s$) region. They should
thus be accurate for $r<r_s$, and could be used as a comparison for other
two-electron approaches to the jellium model, and for testing
reduced-density matrix energy functionals. In particular, it would be
of great interest to compare them
with the geminals which will be computed following the approach
described by P.~Ziesche in this book.

The main lack of the present treatment is the violation of the
particle-conservation sum rule.
A possible solution to this problem is
to write down
a self-consistent set of equations in which the effective potential is unknwon,
but the ``exact'' pair-density of jellium\cite{PW92,noi} is used to 
generate it.

\newpage

\noindent
{\bf ACKNOWLEDGMENTS}
\vspace{0.3cm}

Financial support from MURST (the Italian Ministry for University, Research and
Technology) through COFIN99 is acknowledged.

\thispagestyle{empty}

\end{document}